\documentclass[twocolumn,preprintnumbers,amsmath,amssymb]{revtex4}

\def\be{\begin{equation}}
\def\ee{\end{equation}}

\def\be{\begin{equation}}
\def\ee{\end{equation}}

\def\be{\begin{equation}}
\def\ee{\end{equation}}

\usepackage[dvipdfmx]{graphicx}
\usepackage{here,braket,ulem}
\usepackage{comment}
\usepackage{epsf}
\usepackage{amsmath}
\usepackage{graphics}
\usepackage{amsfonts}
\usepackage{amssymb}
\usepackage{latexsym}
\usepackage{color}
\usepackage{feynmp}

\begin{document}
\preprint{RESCEU-25/16}

\title{Resolution to the firewall paradox:\\The black hole information paradox and highly squeezed interior quantum fluctuations}

\author{Naritaka Oshita}
\affiliation{
  Research Center for the Early Universe (RESCEU), Graduate School
  of Science,\\ The University of Tokyo, Tokyo 113-0033, Japan
}
\affiliation{
  Department of Physics, Graduate School of Science,\\ The University of Tokyo, Tokyo 113-0033, Japan
}


\begin{abstract}
Almheiri, Marolf, Polchinski, and Sully argued that, for a consistent black hole evaporation process, the horizon of a sufficiently old black hole should be replaced by
a ``firewall" at which an infalling observer burns up, which obviously leads to the violation of the equivalence principle.
We propose that once the infalling partner of an outgoing Hawking particle approaches a black hole singularity, it experiences
decoherence and the loss of its entanglement with the outgoing Hawking particle.
This implies we would no longer need firewalls to avoid the black hole information paradox.
\end{abstract}


\maketitle

\section{introduction}
The black hole information paradox \cite{Hawking:1976ra} is one of the most profound problems
in physics, which might lead to a deeper understanding of the relation between
general relativity and quantum theory.
If radiation from a black hole is thermal, the final state of black hole should be a mixed state even
for a black hole originating from a gravitationally collapsing pure state.
This process is forbidden in unitarity of quantum mechanics.
Therefore, it is expected that the radiation from a black hole would be non-thermal.

Susskind, Thorlacius and Ugrum proposed the \textit{black hole complementarity principle} \cite{Susskind:1993if,tHooft:1990fkf,Susskind:1993mu},
which gives the phenomenological picture for the evaporation of black hole that
explains how the non-thermal radiation could be emitted from a black hole.
This proposal is consistent with three postulates, which are briefly given as follows:
(postulate 1) Hawking radiation is in a pure state,
(postulate 2) outside the region near the horizon of a massive black hole,
physics can be described by an effective field theory of general relativity plus quantum field theory,
and (postulate 3) a black hole is regarded as a quantum system with discrete energy levels whose number
is the exponential of the Bekenstein entropy \cite{Bekenstein:1974ax} of black hole.

In 2012, however, Almheiri, Marolf, Polchinski and Sully (AMPS) pointed out in
Ref. \cite{Almheiri:2012rt} that
postulate 1, postulate 2, and the equivalence principle are mutually inconsistent
for an \textit{old black hole} \cite{Page:1993df, Page:1993wv, Page:2013dx} and idea that we briefly review here.
Let us consider an old black hole with early Hawking radiation A, late Hawking radiation B and infalling quanta
behind the horizon C.
A and B have to be fully entangled so that the final state of the black hole is a pure state (postulate 1).
On the other hand, according to quantum field theory in curved spacetime, B and C, pair-created particles, are also fully entangled (postulate 2).
That is, according to postulate 1 and 2, B should be fully entangled simultaneously with both A and C.
This contradicts with the {\it monogamy of entanglement}
that forbids any quantum system being entangled with two independent systems fully and simultaneously.
AMPS then proposed ``{\it firewalls}", high-energy quanta at horizons
energetic enough to break the entanglement of Hawking pairs, which would get rid of the inconsistency between postulate 1 and 2.
However, the existence of firewalls implies that the free falling observer going across the horizon
has a dramatic experience: the observer burns up at the horizon.
That is, firewalls amounts to abandoning the equivalence principle.
\begin{figure}[t]
\begin{center}
\includegraphics[keepaspectratio=true,height=50mm]{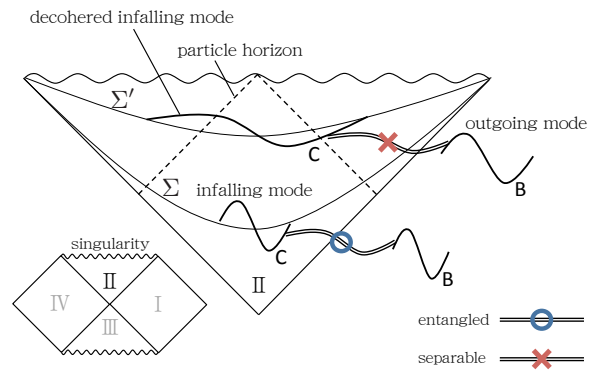}
\end{center}
\caption{
The infalling mode near the horizon, C on the hyper surface $\Sigma$, can hold coherence,
whereas the infalling mode in the vicinity of the singularity, C on the hyper surface $\Sigma'$,
exits the particle horizon (dashed line) and loses causal contact as a whole, which leads to the decoherence of the infalling mode.
As a result, the entanglement of the Hawking pairs disappears and its state becomes separable.
  }%
  \label{061001fig}
\end{figure}

In this paper a sufficient reason for rejecting the AMPS firewall
concept as a solution to the black hole information paradox is
presented. It was previously pointed out that the black hole
information paradox only manifests limitations of the semiclassical
theory, rather than presents a conflict between any fundamental
principles \cite{Nomura:2016qum}. It was proved that firewalls are excluded by
Einstein's field equations for black holes of mass exceeding the
Planck mass \cite{Abramowicz:2013dla}, and demonstrated that the AMPS argument is based on
an over-counting of internal black hole states including those that
are singular in the past \cite{Page:2013mqa}.
Here we show that an infalling mode inside a black hole C is infinitely squeezed due to the gravitational effect of a black hole,
which makes the infalling mode highly sensitive to decoherence
\cite{footnote1}
and leads to the loss of its entanglement with the
outgoing mode B (Fig. \ref{061001fig}).
This means that there would be no violation of monogamy of entanglement around a black hole and
the black hole complementarity principle can be consistent with the equivalence principle.

The plan of this paper is as follows. In Sec. II we will introduce a quantum state around a
black hole formed from gravitational collapse and will describe how we calculate the
time evolution of the quantum state.
The resolution to the firewall paradox is described in Sec. III.
We will show that the quantum state of a Hawking pair, which is initially entangled state,
would become a separable state due to environment-induced decoherence (see e.g.,
\cite{Kieferdeco2} for the review of decoherence).
In Sec. IV we will confirm the consistency between our proposal, explaining how the
Hawking pair evolves to a separable state from an initially entangled state, and the previous works that investigated how the
purity of the Hawking radiation will be realized.
Sec. V is dedicated to conclusions.

\section{formalism}
The Unruh vacuum state \cite{Unruh:1976db} is the quantum state on an eternal black hole spacetime which models the late time properties of the \textit{in vacuum}
of a collapsing star, which is denoted by $\ket{\text{in}}$, that contains no Hawking particle at the past infinity.
The Unruh vacuum is associated with the infalling modes and the outgoing modes that are positive frequency with respect to the Killing vector $\partial_t$ and $\partial_T$ respectively,
where $t$ is the Schwarzschild time and $T$ is the Kruskal time.
Introducing the vacuum state $\ket{0}_c$ for the infalling modes and $\ket{0}_b$ for the outgoing modes,
the Unruh vacuum state can be expressed as $\ket{U} = \ket{0}_c \ket{0}_b$, and the relation between the in vacuum state $\ket{\text{in}}$ and the Unruh vacuum state $\ket{U}$ has the form \cite{Brout:1995rd}
\begin{eqnarray}
\ket{\text{in}} \propto \frac{1}{\sqrt{Z_\omega}} \displaystyle \left( \sum_{n=0}^{\infty} e^{-\pi \omega n(\omega)/\kappa} (b^{\dag}_{\omega})^n (c^{\dag}_{\omega})^n / n! \right)\ket{0}_c \ket{0}_b,
\label{060801}
\end{eqnarray}
where $b^{\dag}_{\omega}$ and $c^{\dag}_{\omega}$ are creation operators for the state $\ket{0}_b$ and $\ket{0}_c$ respectively,
$n(\omega)$ is the number of particles with mode $\omega$, $\kappa \equiv (4 G M)^{-1}$ is the surface gravity, and $Z_{\omega} \equiv (1-e^{-\pi \omega/ \kappa})^{-1}$.
In the following, we will use the Unruh vacuum state as a quantum state around a black hole
although modeling the quantum state around the collapsing star with the (outgoing) Kruskal mode
has not been fully successful and may demand us to take into account the technical issues,
e.g., the backscattering effect in the definition of $\ket{0}_c$ and $\ket{0}_b$
\cite{footnote2}.

The relation (\ref{060801}) implies that the infalling modes are fully entangled with the outgoing modes, which is the problematic entanglement and should be broken for the purity of the Hawking
radiation as is pointed out by AMPS \cite{Almheiri:2012rt}.
In the following, we will neglect multi-pair creations
because the states of {\it n}-particles are suppressed by the exponential factor $e^{-\pi \omega n/\kappa}$
and their cumulative contribution to the entanglement entropy (EE) between the infalling and outgoing mode is negligibly small
\cite{footnote3}.
For simplicity and to grasp the essence, we here consider a generically entangled state
\begin{align}
\ket{\text{in}} &\to \sqrt{1-p^2} \ket{0}_c \ket{0}_b + p \ket{1}_{c} \ket{1}_{b},
\label{061203}\\
\ket{1}_c &= \int_{0}^{\infty} d \omega \varphi_c (\omega) \ket{1,\omega}_c,
\label{170712}\\
\ket{1}_b &= \int_{0}^{\infty} d \omega \varphi_b (\omega) \ket{1,\omega}_b,
\end{align}
where $\ket{1, \omega}_{c} \equiv c^{\dag}_{\omega} \ket{0}_c, \ \ket{1, \omega}_{b} \equiv 
b^{\dag}_{\omega} \ket{0}_b$, $p$ is a real number satisfying $0 < |p| < 1/{\sqrt{2}}$,
and $\varphi_c (\omega)$ $(\varphi_b (\omega))$ is a function satisfying $\int d \omega
 |\varphi_c(\omega)|^2 = 1$ $(\int d \omega
 |\varphi_b(\omega)|^2 = 1)$, which ensures that
$\ket{1}_c$ $(\ket{1}_b)$ is a one-particle state of an infalling (outgoing)
localized wave packet \cite{footnote4}.
In the latter part of this paper, we will show that this entanglement is broken by the existence of the singularity,
which is caused by the decoherence of an infalling mode.
An infalling mode inside a black hole is redshifted \cite{footnote5}
as $\lambda = \lambda_0 \sqrt{2GM/r-1}$, where $\lambda_0$ is the initial wavelength,
and it diverges in the limit of $r \to 0$. Therefore, the infalling mode exits the particle horizon
near the singularity and loses causal contact as a whole (Fig. \ref{061001fig}), which is responsible for the squeezing
(EPR-like correlation) of the infalling mode, that has the role to retain its coherent structure
\cite{Kiefer:2006je}, and decoherence as is discussed in Sec. III.

We consider a massless scalar field $\phi$ on the Schwarzschild spacetime with a mass $M$ whose metric is given as
$ds^2 = f(r) dt^2 - f^{-1}(r) dr^2 -r^2 d\Omega_2^2$ with $f(r) \equiv 1-2GM/r$,
 \ where $d\Omega_2^2$ denotes the line element of a two-sphere $d\Omega_2^2 \equiv d\theta^2 +\sin^2{\theta} d\varphi^2$.
Using the tortoise coordinate $r^{\ast} = r + 2GM \ln \left| 1-r/(2GM) \right|$, we can rewrite it as
$
ds^2 = g_{\mu \nu} dx^{\mu} dx^{\nu} \equiv f(r)\left[ dt^2 -dr^{\ast}{}^2 \right] -r^2 d\Omega_2^2.
$
In order to describe the infinite squeezing of an infalling mode, let us investigate the dynamics of the vacuum $\ket{0}_c$ inside the black hole $r < 2GM$.
The action $S$ is given as
\begin{widetext}
\begin{eqnarray}
S = \int d^4x {\mathcal L}
= \frac{1}{2} \int d^4x \sqrt{-g} g^{\mu \nu} \partial_{\mu} \phi \partial_{\nu} \phi
= \frac{1}{2} \int d^2 x \displaystyle \sum_{l,m} \left[ \chi' {}_{lm}^2 -2 \chi_{lm} \chi' {}_{lm} {\mathcal G} + {\mathcal G}^2 \chi^2_{lm}
- \dot{\chi}_{lm}^2 +f(r) \frac{l (l+1)}{r^2} \chi^2_{lm} \right],
\label{060805}
\end{eqnarray}
\end{widetext}
where we decompose the field $\phi$ into partial waves with an angular momentum $l$ as $\phi \equiv \displaystyle \sum_{l,m} \chi_{lm} Y_{lm} /r$,
a prime and a dot denote differentiation with respect to $r^{\ast}$ and $t$ respectively,
and ${\mathcal G} \equiv r'/r$.
From the action (\ref{060805}), the Euler-Lagrange equation can be derived as
\begin{eqnarray}
\left[ \frac{\partial^2}{\partial r^{\ast} {}^2} - \frac{\partial^2}{\partial t^2} -f(r) \left( \frac{2GM}{r^3} + \frac{l(l+1)}{r^2} \right) \right] \chi_{lm} =0.
 \label{060806}
\end{eqnarray}
We find that the mode functions satisfying (\ref{060806}) are almost independent of the angular momentum $l$ in the vicinity of the singularity
because $l(l+1)/r^2$ in (\ref{060806}) can be ignored for $r \ll 2GM$.
We are interested in the behavior of an infalling mode near the singularity, and
therefore, we set $l=0$ and omit the suffixes $(l,m)$ in the following.
The time like coordinate inside the black hole is $r^{\ast}$, therefore, the conjugate momentum
$\pi$ of the field $\chi$ is given as \cite{Yajnik:1997su}
\begin{eqnarray}
\pi \equiv \partial {\mathcal L}/\partial \chi' = \chi' - {\mathcal G} \chi
\label{063001}
\end{eqnarray}
and then the Hamiltonian is
\begin{eqnarray}
H = \int dt \displaystyle \frac{1}{2} \left[ \pi^2 + \dot{\chi}^2 + 2{\mathcal G} \chi \pi \right].
\label{060803}
\end{eqnarray}
We can decompose the field $\chi$ and its conjugate momentum $\pi$ as
\begin{widetext}
\begin{eqnarray}
&&\chi \equiv \int^{+ \infty}_{-\infty} \frac{d \omega}{\sqrt{2 \pi}} \bar{\chi}_{\omega} (r^{\ast}) e^{-i \omega t} + \text{(O.M.)}
\equiv \int^{+ \infty}_{-\infty} \frac{d \omega}{\sqrt{2 \pi}} \left[ c_{\omega} \tilde{\chi}_{\omega} (r^{\ast}) e^{-i\omega t} + c_{\omega}^{\dag} \tilde{\chi}^{\ast}_{\omega} (r^{\ast}) e^{+i \omega t} \right]  \theta(\omega)+\text{(O.M.)},\label{063003} \\
&&\pi \equiv \int^{+ \infty}_{-\infty} \frac{d \omega}{\sqrt{2 \pi}} \bar{\pi}_{\omega} (r^{\ast}) e^{-i \omega t} + \text{(O.M.)}
\equiv -i \int^{+ \infty}_{-\infty} \frac{d \omega}{\sqrt{2 \pi}} \left[ c_{\omega} \tilde{\pi}_{\omega} (r^{\ast})  e^{-i \omega t} - c_{\omega}^{\dag} \tilde{\pi}^{\ast}_{\omega} (r^{\ast})  e^{+i \omega t} \right] \theta(\omega)+ \text{(O.M.)}, \label{063004}
\end{eqnarray}
\end{widetext}
where (O.M.) denotes the outgoing modes and $\theta (\omega)$ is a step function: $\theta (\omega)=1$ for $\omega > 0$
and $\theta (\omega) = 0$ for $\omega < 0$.
The canonical commutation relation is
$[\bar{\chi}_{\omega}, \bar{\pi}^{\dag}_{\omega'}] = i \delta (\omega-\omega')$.
In the following, we will omit the suffix $\omega$
for simplicity.
From (\ref{063001}) and the canonical commutation relation, we obtain the Wronskian condition as
$\left(\tilde{\chi}'^{\ast} \tilde{\chi} - \tilde{\chi}' \tilde{\chi}^{\ast}  \right) =i$.

The third term in (\ref{060803}) is responsible for the squeezing of infalling modes \cite{Polarski:1995jg,Kiefer:2006je,Lesgourgues:1996jc,Kiefer:2008ku,Kiefer:1998qe}, which becomes stronger as $r^{\ast} \to 0$ as is shown later.
To investigate the dynamics of the states $\ket{0}_c$ and $\ket{1,\omega}_c$,
we first derive the wave functions for them,
$\Psi_0[\bar{\chi}]$ and $\Psi_1[\bar{\chi}]$, that satisfy $c \ket{0}_c=0$
and $\ket{1,\omega}_c =c^{\dag} \ket{0}_c$ respectively.
From (\ref{063003}) and (\ref{063004}), we can rewrite the former in the Schr$\ddot{\text{o}}$dinger representation as
$\left[ \bar{\chi} + i {\gamma}^{-1}(r^{\ast}, \omega) \bar{\pi} \right] \ket{0}_c =0$,
where $\gamma (r^{\ast}, \omega) \equiv \tilde{\pi}^{\ast} / \tilde{\chi}^{\ast}$.
Replacing the conjugate momentum $\bar{\pi}$ by $-i \partial/ \partial \bar{\chi}^{\dag}$, we obtain the wave function
$\Psi_0 [\bar{\chi}]$ of the state $\ket{0}_c$ as
\begin{eqnarray}
\Psi_0 [\bar{\chi}] = \sqrt{\frac{2 \gamma_R}{\pi}} \exp{\left[ -\gamma (r^{\ast}, \omega) \bar{\chi} \bar{\chi}^{\dag} \right]},
\label{061201}
\end{eqnarray}
where $\gamma_R \equiv \text{Re}[\gamma (r^{\ast}, \omega)]$.
On the other hand, $\ket{1,\omega}_c$ satisfies $\ket{1,\omega}_c = c^{\dag} \ket{0}_c$, and hence we obtain
$\Psi_1[\bar{\chi}] \propto \left( \bar{\chi} -\gamma^{\ast} {}^{-1} (r^{\ast}) \partial / \partial \bar{\chi}^{\dag} \right) \Psi_0 [\bar{\chi}]$,
which leads to
\begin{eqnarray}
\Psi_1[\bar{\chi}] = \frac{2\gamma_R}{\sqrt{\pi}} \bar{\chi} \exp{\left[ 
- \gamma (r^{\ast}, \omega) \bar{\chi} \bar{\chi}^{\dag} \right]}.
\label{061202}
\end{eqnarray}
The function $\gamma$ can be calculated numerically from (\ref{060806}).

\section{decoherence near a black hole singularity}
In the following we show that
the density matrix $\rho_{co}$ of the quantum state (\ref{061203})
is reduced to a separable \cite{footnote6}
density matrix $\rho_{de}$ due to the decoherence
once the infalling mode reaches the vicinity of the singularity,
namely, $\rho_{co} \to \rho_{de}$ for $r^{\ast} \to 0$.
To this end, we first show that the infalling mode becomes highly squeezed as the mode approaches the singularity, and secondly,
that the squeezed state is highly sensitive to decoherence.
The density matrix $\rho_{co}$ can be written as
\begin{eqnarray}
\rho_{co} \equiv 
(1-p^2) \ket{0}_c \bra{0}_c \otimes \ket{0}_b \bra{0}_b +p^2 \ket{1}_c \bra{1}_c \otimes \ket{1}_b \bra{1}_b&& \nonumber\\
+p\sqrt{1-p^2} \left( \ket{1}_c \bra{0}_c \otimes \ket{1}_b \bra{0}_b+\ket{0}_c \bra{1}_c \otimes \ket{0}_b \bra{1}_b \right),~~~&&
\label{061407}
\end{eqnarray}
and as is shown later, the separable density matrix $\rho_{de}$ is
\begin{eqnarray}
\rho_{de} = (1-p^2) \ket{0}_c \bra{0}_c \otimes \ket{0}_b \bra{0}_b 
+p^2 \ket{1}_c \bra{1}_c \otimes \ket{1}_b \bra{1}_b.\nonumber \\
\label{061204}
\end{eqnarray}
Hence, we will show that the third and fourth
terms in (\ref{061407}) disappear, that is, $\rho_{co} \to \rho_{de}$, as the infalling mode approaches the vicinity of the singularity.

Let us consider the time evolution of the non-diagonal terms of $\rho_{co}$.
Using (\ref{170712}), $\ket{0}_c \bra{1}_c$ and $\ket{1}_c \bra{0}_c$ in the non-diagonal terms
can be decomposed as
\begin{align}
\begin{aligned}
\ket{0}_c \bra{1}_c &= \int d\omega \varphi_c^{\ast} (\omega) \ket{0}_c \bra{1,\omega}_c,\\
\ket{1}_c \bra{0}_c &= \int d\omega \varphi_c (\omega) \ket{1,\omega}_c \bra{0}_c
\label{062817}
\end{aligned}
\end{align}
respectively, and we will show the decay of $\ket{0}_c \bra{1}_c$ and $\ket{1}_c \bra{0}_c$
by calculating the time evolution of $\ket{0}_c \bra{1, \omega}_c$ and $\ket{1, \omega}_c \bra{0}_c$.
$\ket{0}_c \bra{1,\omega}_c$ and $\ket{1,\omega}_c \bra{0}_c$ component of the Wigner function of $\rho_{co}$, $W_{01}^{(c)}$ and $W_{10}^{(c)}$, are given as
\begin{widetext}
\begin{eqnarray}
\begin{split}
&&W_{01}^{(c)} = W_{10}^{(c)} {}^{\ast}
= \int \int \frac{d x_{R} dx_I}{(2 \pi)^2} e^{-i (\bar{\pi}_{R} x_R + \bar{\pi}_{I} x_I)}
\bra{\bar{\chi} -  \frac{x}{2}} \ket{0}_c \bra{1,\omega}_c \ket{\bar{\chi} + \frac{x}{2}}~~~~~~~~~~~~~~~~~~~~~~~~~~~~~ \\
&&= \frac{1}{\pi^2} \left( \sqrt{2 \gamma_{R}} \bar{\chi} - i \sqrt{\frac{2 \gamma_{I}^2}{\gamma_{R}}} (\bar{\chi} + \frac{\bar{\pi}}{2 \gamma_{I}}) \right) \exp{\left[ -2 \gamma_{R} |\bar{\chi}|^2 \right]}
\exp{\left[ -\frac{2 \gamma_{I}^2}{\gamma_{R}} \left| \bar{\chi} + \frac{\bar{\pi}}{2 \gamma_{I}} \right|^2 \right]}, \label{070501}
\end{split}
\end{eqnarray}
\begin{figure}[t]
  \begin{center}
\includegraphics[keepaspectratio=true,height=82mm]{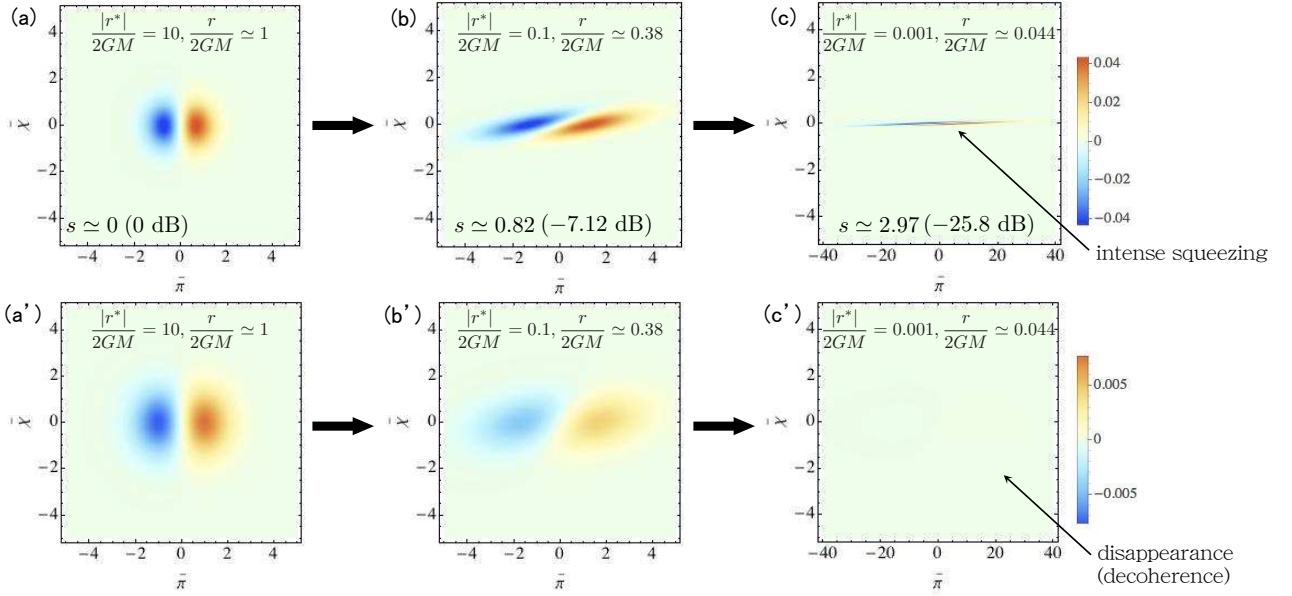}
  \end{center}
  \caption{(a), (b), and (c) are the imaginary parts of the non-diagonal components $W_{01}^{(c)}$,
  and (a'), (b'), and (c') are the imaginary parts of the coarse-grained non-diagonal components
${\mathcal W}_{01}^{(c)}$, where we set
$|r^{\ast}|/2GM = 10$ (for (a), (a')), $|r^{\ast}|/2GM = 0.1$ (for (b), (b')),
$|r^{\ast}|/2GM = 0.001$ (for (c), (c')), and $2GM \omega = 0.5$.
The non-diagonal term
$W_{01}^{(c)} = W_{10}^{(c)}{}^{\ast}$ has the form of $X \delta (X)$ in the limit of $r^{\ast} \to 0$,
and therefore the coarse-grained distribution ${\mathcal W}_{01}^{(c)} = {\mathcal W}_{10}^{(c)} {}^{\ast}$ disappears.
This leads to the transition from the entangled Hawking pair to the separable Hawking pair in the vicinity of the singularity.
  }%
  \label{0630nonfig}
\end{figure}
\end{widetext}
where we used (\ref{061201}) and (\ref{061202}) and the suffixes $R$ and $I$ represent the real and imaginary
part respectively.
We numerically confirmed that they are infinitely squeezed in the limit of $r^{\ast} \to 0$ with $2GM \omega = 0.5$
(Fig.\ref{0630nonfig} (a), (b), and (c))
and the ratio $\gamma_I/\gamma_R \propto \sinh{2s}$ diverges in the vicinity of the singularity, $\gamma_I/\gamma_R \to - \infty$, where $s$ is the squeezing parameter. This means that $s$ also diverges, $|s| \to \infty$, as
$r^{\ast} \to 0$ (see e.g., \cite{Polarski:1995jg}).

Secondly, we will show that an infinitely squeezed state with an environment is highly fragile against decoherence,
in which the environment plays an important role.
For instance, let us consider a double-slit experiment with electrons in which
they create an interference pattern (non-diagonal density matrix).
If they are exposed to thermal noise (environment), the pattern will be coarse-grained
and will disappear (decoherence).
This is the intuitive interpretation for the role of environment in decoherence.
We here take into account the environment as follows.
The field $\chi$ can be
separated into two parts, the long-wavelength part as the system (an infalling Hawking particle) and the short-wavelength part as
the environment (vacuum fluctuations).
We here regard only the modes with wavelengths much shorter than the gravitational
curvature radius of black hole as the short-wavelength part, as in the \textit{stochastic inflation} scheme
\cite{Starobinsky:1994bd,Burgess:2006jn,Calzetta:nqft}.
Therefore, the environment can be regarded as a coherent state with a good approximation
and we can consider the decoherence by tracing out the coherent environment.
It is shown that the tracing out the coherent environment is corresponding to convolving (coarse-graining)
the system's Wigner function (\ref{070501}) with that of a coherent state $W_{\text{E}}$ \cite{Kanada-En'yo:2015pra}
(see also \cite{K.Husimi1940,K.Takahashi1986}),
\begin{eqnarray}
W_{\text{E}} \equiv \pi^{-2} \exp{\left(- |\bar{\chi}|^2- |\bar{\pi}|^2 \right)}.
\label{100501}
\end{eqnarray}
Taking the convolution of (\ref{070501}) and (\ref{100501}), the non-diagonal term of the coarse-grained Wigner function  ${\mathcal W}_{01}^{(c)} = {\mathcal W}_{10}^{(c)} {}^{\ast}$ is obtained as
\begin{widetext}
\begin{eqnarray}
{\mathcal W}_{01}^{(c)} 
\equiv (W^{(c)}_{01} \ast W_E) =\frac{Q |Q|^2}{\pi^2} (\bar{\chi}-i\bar{\pi}) \exp{\left[ -|Q|^2 \left\{ (|\bar{\chi}|^2 + |\bar{\pi}|^2) +2 \gamma_{R}
(|\bar{\chi}|^2 + \left| \bar{\pi}/(2 \gamma_R) + (\gamma_I/\gamma_R) \bar{\chi} \right|^2) \right\} \right]},
\label{070502}
\end{eqnarray}
\end{widetext}
where $Q \equiv \sqrt{2 \gamma_{R}}/(1 + 2 \gamma)$. In the limit of $r^{\ast} \to 0$,
the real and imaginary parts of the function $\gamma (r^{\ast}, \omega)$ diverge and hence $Q$ asymptotically approaches zero.
Therefore, the non-diagonal term ${\mathcal W}_{01}^{(c)}$ is decaying as approaching the
singularity (Fig.\ref{0630nonfig} (a'), (b'), and (c')), which means that the Hawking pair
will experience decoherence as the infalling mode approaches the singularity
since the effect of decoherence on a density matrix is essentially the decay of its non-diagonal
terms, see e.g., \cite{Kieferdeco2}.
Although general relativity and quantum field theory are, of course, no longer valid
near the singularity at $r \lesssim r_{\text{Pl}} = 2GM (M_{\text{Pl}} / M)^{2/3}$
\cite{footnote7}, where $M_{\text{Pl}}$ is the Planck mass,
the decoherence is almost completed
at $r \gg r_{\text{Pl}}$ in the case of interest, namely a massive black hole $M \gg M_{\text{Pl}}$ (remember postulate 2).
That is, the above estimates suggest that the squeezing becomes so strong
that the decoherence can take place well before the modes reach $r \sim r_{\text{Pl}}$,
and therefore using a (semi)classical spacetime picture of the mode evolution should
still be reliable.

As is shown above, the intense squeezing leads to the decay of the non-diagonal terms.
Therefore, the third and fourth terms in (\ref{061407}), containing the non-diagonal components
$\ket{1,\omega}_c \bra{0}_c$ and $\ket{0}_c \bra{1,\omega}_c$ (see (\ref{062817})),
decay due to the decoherence and this leads to the transition of the state
$\rho_{co} \to \rho_{de} = (1-p^2) \ket{0}_c \bra{0}_c \otimes \ket{0}_b \bra{0}_b +
p^2 \ket{1}_c \bra{1}_c \otimes \ket{1}_b \bra{1}_b$. 
This implies that the entanglement of Hawking pairs
decays as the infalling mode approaches the singularity.

\section{microscopic picture of information recovery}
We can apply the loss of the entanglement between a Hawking pair to
the black hole information paradox.
According to our proposal, the entanglement between B and C is broken
when C approaches the singularity. Therefore, the timescale on which the entanglement is broken
is of the order of
the free fall timescale, $t_{F} \sim 2GM$, measured by a freely falling observer
\cite{footnote8}.
In other words, we cannot avoid the entanglement between B and C only during the moment of the free fall $\sim t_{F}$.
Therefore, we have to discuss how the scenario proposed here is consistent with
the monogamy of entanglement and the previous works \cite{Saini:2015dea,Kawai:2015uya}, in which
the timescale of information recovery is carefully discussed in the microscopic level.

\begin{figure}[b]
  \begin{center}
\includegraphics[keepaspectratio=true,height=50mm]{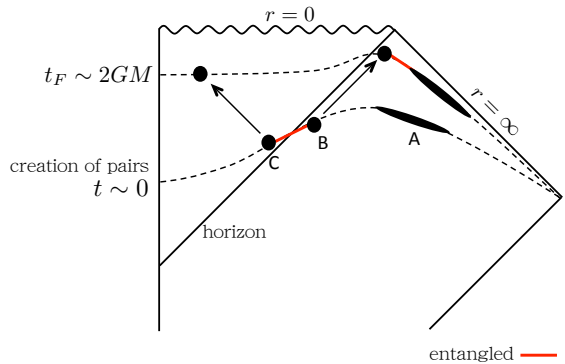}
  \end{center}
  \caption{
The schematic picture showing
how the microscopic picture of information recovery \cite{Saini:2015dea,Kawai:2015uya}
is consistent with our proposal.
B is initially entangled with C and its entanglement will decay on the timescale of $t_{F}$.
On the other hand, the entanglement between A and B is initially zero and may grow on the timescale of $t_{F}$.
  }%
  \label{091102}
\end{figure}

In Ref. \cite{Saini:2015dea}, the radiation around a gravitationally collapsing shell was analytically investigated and it was shown that the correlations between the Hawking
particles (between A and B) are initially zero but grow on the timescale of $t_{F}$ for an observer far from the black hole.
Ref. \cite{Kawai:2015uya} also pointed out that the microscopic timescale of information recovery may be of the order of $t_{F}$
by considering the interaction between a collapsing shell and the Hawking radiation.
For these reasons, we can conclude that the entanglement between A and B would be initially zero and
gradually appears on the timescale of $t_{F}$,
and B can be allowed to be entangled with C only for the short time $ \sim t_F$, which is quite consistent with our scenario.
This implies that B would not be fully entangled with A and C simultaneously (Fig. \ref{091102}), and therefore there is no any violation of the monogamy of entanglement.

\section{conclusions}
We showed that a Hawking pair becomes a separable state from an entangled state
by pointing out that the high squeezing and decoherence occur inside a black hole.
The analysis was done with a simplified state (\ref{061203})
and the environment interacting with the infalling Hawking modes whose Wigner function
is given by (\ref{100501}).
The interaction with the environment can be effectively taken into account by smearing out
the Wigner function of the infalling mode with that of the environment (\ref{070502}).
As a result, we showed that the non-diagonal terms of the density matrix for
the Hawking pair would decay quickly compared to the black hole evaporation timescale,
which implies that the decoherence would be caused by the interior gravitational effect and
that the entanglement between Hawking pairs will be broken.
It should be emphasized that although general relativity and quantum field theory would
break down near the singularity, our proposal is valid as long as the mass of black hole
is much larger than the Planck mass, $M \gg M_{\text{Pl}}$ \cite{Susskind:1993if,Almheiri:2012rt}.
According to our proposal, we would no longer need firewalls.
We believe that our work can be important for the understanding of how the states of
Hawking pairs of particles become separable, and how the black hole information paradox
can be solved.

\section*{acknowledgements}
The author thanks T.~Nakama, Y.~Tada, D.~Yamauchi, and J.~Yokoyama for helpful comments.
This work was partially supported by Grant-in-Aid for JSPS Fellow No.16J01780.


\begin{thebibliography}{99}
\bibitem{footnote3}
The EE is given by $-\displaystyle \sum_{n=0}^{\infty} p_n \ln{p_n}$, where $p_n$ is the probability for $n$-pair creation.
Using Eq. (\ref{060801}) with the typical energy of a Hawking particle $\omega = \kappa$, we can find that the cumulative contribution of the multi-pair creations to the EE is less than $0.1 \%$.
\bibitem{footnote6}
When a density matrix $\rho$ can be rewritten as
$\rho = \displaystyle \sum_{k} p_k \rho^{(c)}_k \otimes \rho^{(b)}_k$ with $\displaystyle \sum_{k} p_k = 1$, the density matrix $\rho$
is said to be ``separable", and this means that there is no entanglement.
\bibitem{footnote7}
The gravitational curvature is of the order of $M_{\text{Pl}}^2$ at $r \sim r_{\text{Pl}}$,
see e.g., \cite{GRA}.
\bibitem{footnote8}
The local velocity is given by $v_{\text{loc}} = ( dr/ dt ) / (1-2GM/r)$.
We can obtain $v_{\text{loc}} = - 1$ and $v_{\text{loc}} = - (2GM/r)^{1/2}$ for a light ray and for a massive particle
falling from rest at infinity respectively \cite{RT1}. Integrating the inverse of local velocity from $r = 2GM$ to $r = 0$, we find out that the free fall time scale
$t_F$ is of the order of $2 G M$ in both cases.
\end{thebibliography}
\end{document}